\documentstyle[floats,aps,preprint]{revtex}
\begin{document}
\draft
\newcommand{\cl}{\centerline}
\renewcommand{\theequation}{\arabic{equation}}
\newcommand\beq{\begin{equation}}
\newcommand\eeq{\end{equation}}
\newcommand\bea{\begin{eqnarray}}
\newcommand\eea{\end{eqnarray}}
%\preprint{\vbox{ \hbox{SOGANG-HEP 263/99}}}
\title{Improved Dirac quantization of a free particle}
\author{Soon-Tae Hong\footnote
{electronic address:sthong@ccs.sogang.ac.kr}, 
        Won Tae Kim\footnote{electronic address:wtkim@ccs.sogang.ac.kr}
   and  Young-Jai Park\footnote{electronic address:yjpark@ccs.sogang.ac.kr}}
\address{Department of Physics and Basic Science Research Institute,\\
         Sogang University, C.P.O. Box 1142, Seoul 100-611, Korea}
\date{\today}
\maketitle
\bigskip 
\begin{abstract}
In the framework of Dirac quantization with 
second class constraints, a free particle moving 
on the surface of a $(d-1)-$dimensional sphere has an
ambiguity in the energy spectrum due to the arbitrary shift of
canonical momenta. We explicitly show that this spectrum obtained by 
the Dirac method can be consistent with 
the result of the Batalin-Fradkin-Tyutin formalism, which is an improved 
Dirac method, at the level of the first-class constraint by fixing
the ambiguity, and discuss its physical consequences.
\end{abstract}
\bigskip

\noindent
\newpage

It is well known that the canonical quantization of a free point particle in
a curved space is a long-standing and controversial problem 
in quantum mechanics \cite{dka,dw,mar,dit}. 
Indeed, for such a system the classical-quantum correspondence
does not uniquely define a Hamiltonian operator and this ambiguity affects 
the energy spectrum of the physical system.

On the other hand, in order to quantize the physical systems subjected to
the constraints, the Dirac quantization scheme \cite{di} has been used
widely. However, whenever we adopt the Dirac method, we frequently meet the problem
of the operator ordering ambiguity. In order to avoid this problem, Batalin,
Fradkin, and Tyutin (BFT) developed a method \cite{BFT} converting the
second-class constraints into first-class ones, which instead of 
configuration space restricts the quantum-mechanical Hilbert space.
Then, the operators representing the first-class constraints 
are generators of gauge transformations, and the physical states are all found 
by going into the gauge invariant subspace of the Hilbert space.
Recently, this BFT formalism has been applied to several interesting
models \cite{BFT1}. Very recently, the SU(2) Skyrme model has been studied
in the context of the BFT formalism \cite{sk2,hkp}.

In this Letter, we will perform a Hamiltonian quantization of 
a free particle moving on the surface of a $(d-1)-$dimensional sphere by
exactly identifying the ambiguity of the energy spectrum. 
We show how this ambiguous spectrum obtained by the Dirac method 
can be consistent with the result of the BFT formalism, 
which is an improved version of Dirac method.
Firstly, the Dirac bracket scheme will be applied to a  
free particle constrained on a $(d-1)-$dimensional sphere. 
The adjustable parameter will be
introduced to define the generalized momenta without any loss of generality,
which yields an ambiguous energy spectrum.
Next, we will apply the BFT method to this model to obtain the
energy spectrum including the Weyl ordering correction.
Then, we will show that by fixing this free parameter 
the energy eigenvalues obtained by the Dirac method are consistent 
with the result of the BFT formalism.  Finally, we will construct 
the BRST invariant gauge fixed Lagrangian 
\cite{brst} as well as the effective Lagrangian 
corresponding to the first-class Hamiltonian in the Batalin, Fradkin and 
Vilkovisky (BFV) scheme \cite{bfv,fik,biz}.

Now we start with the following Lagrangian describing 
a free particle with a unit mass on a $(d-1)-$dimensional 
sphere of unit radius embedded in a $d-$dimensional 
Cartesian space with coordinates 
$q_i(i=1,2,..,d)$:
\begin{equation}
L= \frac{1}{2}\dot{q}_{i}\dot{q}_{i}.
\label{lag}
\end{equation}
Introducing the canonical momenta $\pi_{i}=\dot{q}_{i}$ conjugate 
to the coordinates $q_{i}$ one can then obtain the canonical 
Hamiltonian 
\begin{equation}
H=\frac{1}{2}\pi_{i}\pi_{i}.
\label{hamil}
\end{equation}

On the other hand, we have the following second-class constraints:
$$
%\begin{eqnarray}
\Omega_{1} = q_{i}q_{i}-1\approx 0,~~ 
\Omega_{2} = q_{i}\pi_{i}\approx 0, 
%\label{omega2}
$$
%\end{eqnarray}
to yield the Poisson algebra $\Delta_{ab}=\{\Omega_{a},\Omega_{b}\} = 
2\epsilon^{a b}q_{i}q_{i}$ with $\epsilon^{12}=-\epsilon^{21}=1$. 
Here one notes that, due to the commutator $\{\pi_{i},\Omega_{1}\}
=-2q_i$, one can easily obtain the algebraic relation 
$\{\Omega_1,H\}=2\Omega_2$.  Using the 
Dirac brackets \cite{di} defined by $\{A,B\}_{D}=\{A,B\}-\{A,\Omega_{a}\}
\Delta^{a b}\{\Omega_{b},B\}$ with $\Delta^{a b}$ being the inverse of 
$\Delta_{a b}$ and performing the canonical quantization scheme $\{A,B\}_{D}
\rightarrow \frac{1}{i}[A_{op},B_{op}]$, one can obtain the operator 
commutators 
\begin{eqnarray}
{[q_{i},q_{j}]}&=&0, \nonumber \\
{[q_i,\pi_j]}&=&i(\delta_{i j}-\frac{q_{i}q_{j}}{q_{k}q_{k}}),  \nonumber \\
{[\pi_i,\pi_j]}&=&\frac{i}{q_{k}q_{k}} (q_{j}\pi_{i}-q_{i}\pi_{j})  
\label{aaop}
\end{eqnarray}
with $\pi_{i}=-i(\delta_{ij}-\frac{q_{i}q_{j}}{q_{k}q_{k}})
\partial_{j}$. 

Then we observe that without any loss of generality the generalized momenta $
\Pi_{i}$ fulfilling the structure of the commutators (\ref{aaop}) are given by  
$\Pi_i=-i(\delta_{ij}-\frac{q_{i}q_{j}}{{q_{k}q_{k}}})
\partial_{j}-\frac{ic q_i}{{q_{k}q_{k}}}$ with an arbitrary parameter $c$ to 
be fixed later \cite{hkp,corr}.  In Ref. \cite{no,fujii} the authors did not 
include the last term so that one cannot clarify the relations between the BFT 
scheme and the Dirac bracket one.  On the other hand, the energy spectrum of 
the free particle can be obtained in the Weyl ordering scheme \cite{weyl} 
where the Hamiltonian (\ref{hamil}) is modified into the symmetric form 
$$
%\begin{equation}
H_{N}=\frac{1}{2}\Pi_i^{N}\Pi_i^{N},  \label{hamil2}
$$
%\end{equation}
where 
$$
%\begin{equation}
\Pi_i^{N}=-\frac{i}{2}
\left[(\delta_{ij}-\frac{q_{i}q_{j}}{{q_{k}q_{k}}})\partial_{j}
+\partial_{j}(\delta_{ij}-\frac{q_{i}q_{j}}{{q_{k}q_{k}}})
+\frac{2c q_i}{{q_{k}q_{k}}}
\right].
$$
%\end{equation}
After some algebra, one can obtain the Weyl ordered $\Pi_i^{N}\Pi_i^{N}$ 
as follows:
\begin{equation}
\ \Pi_i^{N}\Pi_i^{N}=-\partial_{i}\partial_{i}
+\frac{(d-1)q_i}{{q_{k}q_{k}}}\partial_{i}
+\frac{q_{i}q_{j}}{{q_{k}q_{k}}}\partial_{i}\partial_{j}
+\frac{1}{q_{k}q_{k}}\left[\frac{(d-1)^2}{4}-c^{2}\right],
\label{eq}
\end{equation}
which yields the modified quantum energy spectrum
as 
\begin{equation}
\langle H_{N}\rangle= \frac{1}{2}
\left[l(l+d-2)+\frac{(d-1)^2}{4}-c^{2}\right].
\label{hwc}
\end{equation}
Here the first three terms in Eq. (\ref{eq})
are nothing but the $(d-1)-$sphere Laplacian \cite
{Vil} given in terms of the coordinates and their derivatives to
yield the eigenvalues $l(l+d-2)$.
Note that due to the ambiguity of the arbitrary value $c$, 
we could adjust any energy spectrum 
obtained by various approaches\cite{dw,dit} to give the proper spectrum. 
In fact one cannot fix uniquely 
the energy spectrum only by using the Dirac method.

Next, following the Abelian BFT formalism \cite{BFT,BFT1,sk2} which 
systematically converts the second-class constraints into the first-class ones,
we introduce two auxiliary fields $\Phi^{a}$ corresponding to $\Omega_{a}$
with the Poisson brackets $\{\Phi^{a}, \Phi^{b}\}=\omega^{ab}$.  The 
first-class constraints $\tilde{\Omega}_{a}$ are then constructed as a 
power series of the auxiliary fields: 
\begin{equation}
\tilde{\Omega}_{a}=\sum_{n=0}^{\infty}\Omega_{a}^{(n)},~~~~
\Omega_{a}^{(0)}=\Omega_{a}  \label{tilin}
\end{equation}
where $\Omega_{a}^{(n)}$ are polynomials in the auxiliary fields $\Phi^{a}$
of degree $n$, to be determined by the requirement that the first-class
constraints $\tilde{\Omega}_{a}$ satisfy an Abelian algebra $\{\tilde
{\Omega}_{a},\tilde{\Omega}_{b}\}=0$.  Following the standard iterating 
procedure with the choice of $\omega^{ab}=\epsilon^{ab}$
\cite{BFT,BFT1}, one 
can obtain the first-class constraints 
\beq
\tilde{\Omega}_{1}=\Omega_{1}+2\Phi^{1},~ 
\tilde{\Omega}_{2}=\Omega_{2}-q_{i}q_{i}\Phi^{2},  
\label{1stconst}
\eeq
which yield the strongly involutive first-class constraint algebra $\{\tilde
{\Omega}_{a},\tilde{\Omega}_{b}\}=0$. 
%Since $\Omega_{a}^{(1)}$ are linear in the auxiliary fields, one can make
%the ansatz 
%\begin{equation}
%\Omega_{a}^{(1)}=X_{ab}\Phi^{b}.  \label{xijphi}
%\end{equation}
%Substituting Eq. (\ref{xijphi}) into Eq. (\ref{cijk}) leads to the following
%relation:
%$$
%\begin{equation}
%\Delta_{ab}+X_{ac}\omega^{cd}X_{bd}=0,  
%$$
%\label{delx}
%\end{equation}
%which, for the standard choice \cite{BFT,BFT1} of $\omega^{ab}=
%\epsilon^{ab}$, has a solution 
%\begin{equation}
%X_{ab}=\left( 
%\begin{array}{cc}
%2 & 0 \\ 
%0 & -q_{i}q_{i}
%\end{array}
%\right).  \label{xij}
%\end{equation}
%Substituting Eq. (\ref{xij}) into Eqs. (\ref{tilin}) and (\ref{xijphi}) and 
%iterating this procedure, 

%%%%%%%%%%%%%%%%%%%%%%%%%%
Now we systematically construct the first-class BFT physical fields 
$\tilde{{\cal F}}=(\tilde{q}_i, \tilde{\pi}_i)$ in the extended phase space
corresponding to the original fields ${\cal F}=(q_i,\pi_i)$, which are
obtained as a power series in the auxiliary fields $\Phi^{a}$ by demanding
that they are strongly involutive: $\{\tilde{\Omega}_{a}, \tilde{{\cal F}}
\}=0$. In general the first-class-fields satisfying the boundary conditions 
$\tilde{{\cal F}}[{\cal F};0]={\cal F}$ can be found as 
$$
%\begin{equation}
\tilde{{\cal F}}[{\cal F};\Phi]={\cal F}+\sum_{n=1}^{\infty} \tilde{%
{\cal F}}^{(n)},~~~ \tilde{{\cal F}}^{(n)}\sim (\Phi)^{n}
$$
%\end{equation}
where the $(n+1)-$th order iteration terms are given by the formula 
$$
%\begin{equation}
\tilde{{\cal F}}^{(n+1)}=-\frac{1}{n+1}\Phi^{a}\omega_{ab}X^{bc}G_{c}^{(n)}
$$
%\end{equation}
with 
$$
%\begin{equation}
G_{a}^{(n)}=\sum_{m=0}^{n}\{\Omega_{a}^{(n-m)},\tilde{{\cal F}}^{(m)} \}_{(%
{\cal F)}}+\sum_{m=0}^{n-2}\{\Omega_{a}^{(n-m)},\tilde{{\cal F}}^{(m+2)}
\}_{(\Phi)}+\{\Omega_{a}^{(n+1)},\tilde{{\cal F}}^{(1)}\}_{(\Phi)}.
$$
%\end{equation}
After some algebra, we obtain the first-class physical fields,
\begin{eqnarray}
\tilde{q}_i&=&q_i\left[1-\sum_{n=1}^{\infty}\frac{(-1)^{n}(2n-3)!!}{n!} 
\frac{(\Phi^{1})^{n}}{(q_{k}q_k)^{n}}\right]  \nonumber \\
\tilde{\pi}_i&=&(\pi_{i}-q_{i}\Phi^{2})\left[1+\sum_{n=1}^{\infty}
\frac{(-1)^{n}(2n-1)!!}{n!}\frac{(\Phi^{1})^{n}}{(q_{k}q_k)^{n}}\right]
\label{pitilde}
\end{eqnarray}
with $(-1)!!=1$.

Then, using the novel property\cite{kkppy} that any functional ${\cal K}(%
\tilde{{\cal F}})$ of the first-class fields $\tilde{{\cal F}}$ will also be
first-class, i.e., $\tilde{{\cal K}}({\cal F};\Phi )={\cal K}
(\tilde{{\cal F}})$, we can directly construct the first-class Hamiltonian in 
terms of the above BFT physical variables as follows 
$$
%\begin{equation}
\tilde{H}= \frac{1}{2}\tilde{\pi}_i\tilde{\pi}_i
$$
%\label{htilde}
%\end{equation}
omitting infinitely iterated standard procedure \cite{hkp}. 

As a result, the corresponding first-class Hamiltonian 
with the original fields and auxiliary fields is given by 
\begin{equation}
\tilde{H}= \frac{1}{2}(\pi_i-q_i\Phi^{2})
(\pi_i-q_i\Phi^{2})\frac{q_{j}q_{j}}{q_{j}q_{j}+2\Phi^{1}},
\label{hct}
\end{equation}
which is also strongly involutive with the first-class constraints $\{\tilde
{\Omega}_{a},\tilde{H}\}=0$.  However, with the Hamiltonian (\ref{hct}), one 
cannot naturally generate the first-class Gauss' law constraint from the time 
evolution of the primary constraint $\tilde{\Omega}_{1}$. By introducing an
additional term proportional to the first-class constraints $\tilde{\Omega}%
_{2}$ into $\tilde{H}$, we obtain an equivalent first-class Hamiltonian 
\begin{equation}
\tilde{H}^{\prime}=\tilde{H}+\Phi^{2}\tilde{\Omega}_{2},
\label{hctp}
\end{equation}
which naturally generates the Gauss' law constraint, $\{\tilde{\Omega}_{1},
\tilde{H}^{\prime}\}= 2\tilde{\Omega}_{2}$ and $\{\tilde{\Omega}_{2},\tilde{H}
^{\prime}\}=0$.  Here one notes that $\tilde{H}$ and $\tilde{H}^{\prime}$ act 
on physical 
states in the same way since such states are annihilated by the first-class 
constraints. Similarly, the equations of motion for observables are also 
unaffected by this difference. Furthermore, if we take the limit 
$\Phi^{a}\rightarrow 0$, then our first-class system exactly returns to the 
original second-class one. 
%On the other hand, 
%using the first-class constraints in the Hamiltonian (\ref{hctp}), 
%one can obtain a Hamiltonian of the form 
%$$
%\begin{equation}
%\tilde{H}^{\prime}= \frac{1}{2}(q_iq_i\pi_{j}\pi_{j}
%-q_i\pi_{i}q_j\pi_{j}).  
%$$
%\label{htilde2}
%\end{equation}

We are now ready to obtain the energy spectrum of the extended phase
space. The fundamental idea consists in imposing quantum mechanically
the first-class constraints as operator condition on the state as a way 
to obtain the physical subspace, $i.e.$, $\tilde{\Omega}_{a}| {\rm phys}>=0$ 
where we used the symmetrized operators as 
$\tilde{\Omega}_{1}=q_i q_i + 2 \Phi^1 $ and $\tilde{\Omega}_{2}
=(q_i \pi_i)_{{\rm sym}} -q_i q_i \Phi^2 $.
Then, after the symmetrization procedure\cite{sk2}, 
the first-class Hamiltonian yields
the energy spectrum with the Weyl ordering correction  
$$
%\begin{equation}
\langle\tilde{H}^{\prime}_{N}\rangle= \frac{1}{2}\left[
 l(l+d-2)+\frac{d(d-3)}{4}\right].
$$
%\label{nht}
%\end{equation}
This result obtained through the Abelian BFT analysis is well in 
agreement with the energy level spacings due to the angular contribution of 
the hydrogen atom because there is no additional constant parameter 
in the energy eigenvalues for the case of $d=3$. 
Furthermore, our result well describes the spectrum of SU(2) 
Skyrmion model 
corresponding to the $d=4$ case \cite{sk2,hkp}.
Note that, however, the recent result obtained from
the unusual non-Abelian BFT scheme \cite{no} can not describe
the correct situation for $d=3$ case. 
The reason is that this can not naturally generate the Gauss' law constraint, 
and does not recover the original second-class constraint structure
in the limit of $\Phi^a \rightarrow 0$.

Now, in order for the Dirac bracket scheme to be consistent with the BFT
one, the adjustable parameter $c$ in Eq. (\ref{hwc}) should be fixed with the
values 
$$
%\begin{equation}
c=\pm\frac{\sqrt{d+1}}{2}.
$$
%\end{equation}
Then, this fixed parameter $c$ relates the Dirac 
bracket scheme to the BFT one to yield the desired quantization in the 
model of the free particle on $(d-1)-$sphere so that one can achieve 
the unification of these two formalisms.
%%%%%%%%%%%%%%%%%%%%%%%%%%%%%%%%%%%%%%%%%%%%%%%%%%%%%%%%%%%%%%%%%%%%%%

Next, let us consider the partition function of the model in order to present
the Lagrangian corresponding to the first-class Hamiltonian $\tilde{H}%
^{\prime}$ in the canonical Hamiltonian formalism. First of all,
let us identify
the auxiliary fields $\Phi^{a}$ with a canonical conjugate pair $%
(\theta,\pi_{\theta})$, i.e., $\Phi^{a}=(\theta,\pi_{\theta})$ which satisfy 
$\{\Phi^{a}, \Phi^{b}\}=\omega^{ab}$ with $\omega^{ab}=\epsilon^{ab}$. 
Then, the starting partition function in the phase space is given by 
the Faddeev-Senjanovic formula \cite{fads} as follows 
$$
%\begin{eqnarray}
Z=N\int {\cal D}q_i{\cal D}\pi_i{\cal D}\theta{\cal D}\pi_{\theta}
\prod_{a,b=1}^{2}\delta(\tilde{\Omega}_a)\delta(\Gamma_b)\det|M|
\exp i\int {\rm d}t (\pi_{i}\dot{q}_{i}+\pi_{\theta}\dot{\theta}
-\tilde{H}^{\prime})
$$
%\nonumber
%\end{eqnarray}
where the gauge fixing conditions $\Gamma_{i}$ are chosen so that the
determinant occurring in the functional measure is nonvanishing, and $M=\{\tilde
{\Omega}_a,\Gamma_b\}$.

Now, exponentiating the delta function $\delta (\tilde{\Omega}_{2})$ as 
$\delta (\tilde{\Omega}_{2})=\int {\cal D}\xi e^{i\int {\rm d}t~\xi 
\tilde{\Omega}_{2}}$ and performing the integration over $\pi_{\theta}$, we 
obtain 
\begin{eqnarray}
Z&=& N\int {\cal D}q_i{\cal D}\pi_i{\cal D}\theta{\cal D}\xi
\delta(q_{i}q_i-1+2\theta)\prod_{a=1}^{2}\delta(\Gamma_a)\det|M|e^{i\int {\rm d}t L}  \nonumber \\
L&=& -\frac{1}{2}q_{i}q_i\pi_{j}\pi_j +(\dot{q}_{i} -\xi q_i)\pi_i-\frac{1}
{2(q_{k}q_k)^{2}}(\dot{\theta}+\xi q_{i}q_i)^{2}.\nonumber
\end{eqnarray}
After integrating out the momenta $\pi_i$ and auxiliary field $\xi$, 
the partition function is given as follows
\begin{eqnarray}
Z&=&N\int {\cal D}q_i{\cal D}\theta\delta(q_{i}q_i-1+2\theta)
\prod_{a=1}^{2}\delta(\Gamma_a)\det|M| e^{i\int {\rm d}t L}\label{fca}\\
L&=& \frac{1}{2q_{k}q_k} \dot{q}_{i}\dot{q}_{i}-%
\frac{1}{2(q_{k}q_k)^{2}} \dot{\theta}^{2}.  \label{zhct}
\end{eqnarray}
As a result, we have obtained the desired Lagrangian (\ref{zhct}) corresponding
to the first-class Hamiltonian (\ref{hctp}).  Here one notes that the 
Lagrangian (\ref{zhct}) can be re-shuffled to yield the gauge invariant action 
of the form 
\begin{eqnarray}
S&=& \int {\rm d}t \ (\frac{1}{2}\dot{q}_{i}\dot{q}_i) +S_{WZ}  \nonumber
\\
S_{WZ}&=&\int {\rm d}t \left[ \frac{1}{q_{k}q_k} 
\dot{q}_{i}\dot{q}_i\theta -\frac{1}{2(q_{k}q_k)^2} {\dot{\theta}}^{2} \right],
\nonumber
\end{eqnarray}
where $S_{WZ}$ is the new type of the Wess-Zumino term restoring the gauge
symmetry under the transformation: $\delta q_i = q_i \epsilon,~\delta \theta 
= - q_{i}q_i\epsilon$ where $\epsilon$ is a local gauge parameter.
Here one notes that this form of symmetry 
transformation is exactly the same as that 
obtained when we consider the effective first-class constraints 
(\ref{1stconst}) as the symmetry generators in the Hamiltonian formalism.

Moreover the corresponding partition function (\ref{fca}) can be
rewritten simply in terms of the first-class physical fields (\ref{pitilde}) 
$$
%\begin{eqnarray}
\tilde{Z}=N\int {\cal D}\tilde{q}_{i}\delta(\tilde{q}_{j}
\tilde{q}_{j}-1)\prod_{a=1}^{2}\delta(\Gamma_a)\det|M|
\exp i\int {\rm d}t (\frac{1}{2}\dot{\tilde{q}}_i\dot{\tilde{q}}_i)  
$$
%\label{zhct2}
%\end{eqnarray}
where $\tilde{L}$ is form invariant Lagrangian of Eq. (\ref{lag}).

%%%%%%%%%%%%%%%%%%%%%%%%%%%%%%%%%%%%%%%%%%%%%%%%%%%%%%%%%%%%%%%%%%%%%%%%%
Now, in order to obtain the BRST invariant gauge fixed Lagrangian, we
introduce two canonical sets of ghosts and anti-ghosts together with
auxiliary fields in the framework of the BFV formalism \cite{bfv,fik,biz},
which is applicable to theories with the first-class constraints: $({\cal C}
^{a},\bar{{\cal P}}_{a}),~({\cal P}^{a},\bar{{\cal C}}_{a}),~(N^{a},B_{a}),
~(a=1,2)$ which satisfy the super-Poisson algebra
 $\{{\cal C}^{a},\bar{{\cal P}}_{b}\}
=\{{\cal P}^{a},\bar{{\cal C}}_{b}\}=\{N^{a},B_{b}\}=\delta_{b}^{a}$.
The super-Poisson bracket is defined as $\{A,B\}=\frac{\delta A}
{\delta q}|_{r}\frac{\delta B}{\delta p}|_{l}-(-1)^{\eta_{A}\eta_{B}}\frac
{\delta B}{\delta q}|_{r}\frac{\delta A}{\delta p}|_{l}$ where $\eta_{A}$ 
denotes the number of fermions called ghost number in $A$ and the subscript 
$r$ and $l$ right and left derivatives.
In the model for the free particle on a $(d-1)-$dimensional sphere, the 
nilpotent BRST charge $Q$, the fermionic gauge fixing function $\Psi$ and the 
BRST invariant minimal Hamiltonian $H_{m}$ are given by
$$
%\begin{eqnarray}
Q={\cal C}^{a}\tilde{\Omega}_{a}+{\cal P}^{a}B_{a},~~
\Psi=\bar{{\cal C}}_{a}\chi^{a}+\bar{{\cal P}}_{a}N^{a},~~  
H_{m}=\tilde{H}^{\prime}-2{\cal C}^{1}\bar{{\cal P}}_{2}
$$
%\end{eqnarray}
which satisfy the relations $\{Q,H_{m}\}=0,~Q^{2}=\{Q,Q\}=0$, and 
$\{\{\Psi,Q\},Q\}=0$.  The effective quantum Lagrangian is then described as
$$
%\begin{equation}
L_{eff}=\pi_{i}\dot{q}_{i}+\pi_{\theta}\dot{\theta}+B_{2}\dot{N}^{2}+%
\bar{{\cal P}}_{a}\dot{{\cal C}}^{a}+\bar{{\cal C}}_{2} \dot{{\cal P}}%
^{2}-H_{tot}
$$
%\end{equation}
with $H_{tot}=H_{m}-\{Q,\Psi\}$. Here $B_{1}\dot{N}^{1} +\bar{{\cal C}}_{1}%
\dot{{\cal P}}^{1}=\{Q,\bar{{\cal C}}_{1} \dot{N}^{1}\}$ terms are
suppressed by replacing $\chi^{1}$ with $\chi^{1} +\dot{N}^{1}$.

Now we choose the unitary gauge $\chi^{1}=\Omega_{1},~\chi^{2}=\Omega_{2}$ and 
perform the path integration over the fields $B_{1}$, $N^{1}$, $\bar{%
{\cal C}}_{1}$, ${\cal P}^{1}$, $\bar{{\cal P}}_{1}$ and ${\cal C}^{1}$, by
using the equations of motion, to yield the effective Lagrangian of the form
\begin{eqnarray}
L_{eff}&=&\pi_{i}\dot{q}_{i}+\pi_{\theta}\dot{\theta} +B\dot{N}+\bar{%
{\cal P}}\dot{{\cal C}}+\bar{{\cal C}}\dot{{\cal P}}  \nonumber \\
& &-\frac{1}{2}(\pi_{i}-q_{i}\pi_{\theta})(\pi_{i}-q_{i}
\pi_{\theta})\frac{q_{j}q_{j}}{q_{j}q_{j}+2\theta} -%
\pi_{\theta}\tilde{\Omega}_{2}  \nonumber \\
& &+2q_{i}q_{i}\pi_{\theta}\bar{{\cal C}}{\cal C}+\tilde{\Omega}_{2}N
+B\Omega_{2}+\bar{{\cal P}}{\cal P}\nonumber
\end{eqnarray}
with redefinitions: $N\equiv N^{2}$, $B\equiv B_{2}$, $\bar{{\cal C}}\equiv
\bar{{\cal C}}_{2}$, ${\cal C}\equiv {\cal C}^{2}$, $\bar{{\cal P}}\equiv
\bar{{\cal P}}_{2}$, ${\cal P}\equiv {\cal P}_{2}$.

Then, using the variations with respect to $\pi_{i}$, $\pi_{\theta}$, 
${\cal P}$ and $\bar{{\cal P}}$ and identifying $N$ with $N=-B+\frac{\dot
{\theta}}{q_{k}q_{k}}$, we obtain the desired effective Lagrangian
%\begin{eqnarray}
%\dot{q}_{i}&=&(\pi_{i}-q_{i}\pi_{\theta})
%q_{k}q_{k}+q_{i}(\pi_{\theta}-N-B)  \nonumber\\
%\dot{\theta}&=&-q_{i}(\pi_{i}-q_{i}\pi_{\theta})
%q_{k}q_{k} +q_{i}q_{i}(-2\pi_{\theta}-2\bar{%
%{\cal C}}{\cal C}+N) +q_{i}\pi_{i}  \nonumber \\
%{\cal P}&=&-\dot{{\cal C}},~~~~~\bar{{\cal P}}=\dot{\bar{{\cal C}}}
%\nonumber
%\end{eqnarray}
%\begin{eqnarray}
%L_{eff}&=&\frac{1}{2q_{k}q_{k}}\dot{q}_{i}\dot{q}%
%_{i}-\frac{1}{2}\left[\frac{\dot{\theta}}{q_{k}q_{k}} +(B+2\bar{%
%{\cal C}}{\cal C})q_{k}q_{k}\right]^{2}  
%+B\dot{N}+\dot{\bar{{\cal C}}}\dot{{\cal C}}\nonumber\\
%& &+\frac{q_{i}}{q_{k}q_{k}}\left[ \dot{q}%
%_{i}+q_{i}(\frac{\dot{\theta}} {q_{k}q_{k}}+(B+2\bar{{\cal C}}%
%{\cal C})q_{k}q_{k})\right] (B+N).  \nonumber 
%\end{eqnarray}
$$
%\begin{eqnarray}
L_{eff}=\frac{1}{2q_{k}q_{k}}\dot{q}_{i}\dot{q}_{i}
-\frac{1}{2(q_{k}q_{k})^{2}}\dot{\theta}^{2}
-\frac{1}{2}(q_{k}q_{k})^{2}(B+2\bar{{\cal C}}{\cal C})^{2}  
-\frac{\dot{\theta}\dot{B}}{q_{k}q_{k}} +\dot{\bar{{\cal C}}}%
\dot{{\cal C}},
$$
%\nonumber
%\end{eqnarray}
which is invariant under the BRST transformation
\begin{eqnarray}
\delta_{B}q_{i}&=&\lambda q_{i}{\cal C},~~~ \delta_{B}\theta=-\lambda
q_{_i}q_{i}{\cal C},  \nonumber \\
\delta_{B}\bar{{\cal C}}&=&-\lambda B,~~~ \delta_{B}{\cal C}=\delta_{B}B=0.
\nonumber
\end{eqnarray}
Here one notes that the above BRST transformation including the rules for
the (anti)ghost fields is just the generalization of the previous one 
$\delta q_i = q_i \epsilon,~\delta \theta = - q_{i}q_i\epsilon$.

%%%%%%%%%%%%%%%%%%%%%%%%%%%%%%%%%%%%%%%%%%%%%%%%%%%%%%%%%%%%%%%%%%%%%%%%%%%%

In summary, we have clarified the relation between the Dirac bracket scheme
with the second-class constraints
and the BFT method at the level of the first-class constraint, 
which has been obscure and unsettled, in the 
free particle on a $(d-1)-$dimensional sphere. In this approach 
we have introduced the generalized
momentum operators including the free parameter, which is fixed to yield the
consistency between these two formalisms.  We have shown that one could 
see the effects of the Weyl ordering correction in the energy spectrum.  
Note that the energy spectrum is remarkably reproduced 
for $d=3$ and $d=4$ which correspond to the explicit physical phenomena,
three-dimensional rotator and SU(2) Skyrmion, respectively.

%\vskip 1.0cm
{\bf Acknowledgments}\\

The present work was supported by the Sogang University Research Grants in 
1999 and by the Ministry of Education, BK21 Project No. D-1099, 1999.

%%%%%%%%%%%%%%%%%%%% References %%%%%%%%%%%%%%%%%%%%%%%%%

\end{document}